\documentclass[12pt]{article}

\usepackage{hyperref,bm,amsmath,amssymb,a4wide,graphicx,subfigure,wasysym,amsfonts,cite}

\begin{document}

\title{\textbf{Magnetic field instability driven by anomalous magnetic moments of massive fermions and electroweak interaction with background matter}}

\author{Maxim Dvornikov}

\author{Maxim Dvornikov$^{a,b}$\thanks{maxdvo@izmiran.ru}
\\
$^{a}$\small{\ Pushkov Institute of Terrestrial Magnetism, Ionosphere} \\
\small{and Radiowave Propagation (IZMIRAN),} \\
\small{108840 Troitsk, Moscow, Russia;} \\
$^{b}$\small{\ Physics Faculty, National Research Tomsk State University,} \\
\small{36 Lenin Avenue, 634050 Tomsk, Russia}}

\date{}

\maketitle

\begin{abstract}
It is shown that the electric current of massive fermions along the external magnetic field can be excited in the case when particles possess anomalous magnetic moments and electroweakly interact with background matter. This current is calculated on the basis of the exact solution of the Dirac equation in the external fields. It is shown that the magnetic field becomes unstable if this current is taken into account in the Maxwell equations. Considering a particular case of a degenerate electron gas, which can be found in a neutron star, it is revealed that the seed magnetic field can be significantly enhanced. The application of the results to astrophysics is also discussed.
\end{abstract}

%%%%%%%%%%%%%%%%%%%%%%%%%%%%%%%%%%

The problem of the magnetic field instability is important, e.g.,
in the context of the existence of strong astrophysical magnetic fields~\cite{Spr08}.
Besides the conventional magnetohydrodynamics mechanisms for the generation of
astrophysical magnetic fields, recently the approaches based on the
elementary particle physics were proposed. These approaches mainly
rely on the chiral magnetic effect (CME)~\cite{Vil80}, which
consists in the appearance of the anomalous current of massless charged
particles along the external magnetic field $\mathbf{B}$,
\begin{equation}\label{eq:CME}
  \mathbf{J}_\mathrm{CME} = \frac{e^2}{4\pi^2}
  \left(
    \mu_{\mathrm{R}} -\mu_{\mathrm{L}}
  \right)\mathbf{B},
\end{equation}
where $e$ is the particles charge and $\mu_{\mathrm{R,L}}$ are the chemical potentials of right and
left chiral fermions.

If $\mathbf{J}_\mathrm{CME}$ is accounted for in the
Maxwell equations, a seed magnetic field appears to be unstable and can
experience a significant enhancement. The applications of CME for
the generation of astrophysical and cosmological magnetic fields are
reviewed in Ref.~\cite{Sig17}.

However, the existence of CME in astrophysical media is questionable.
As found in Refs.~\cite{Vil80,Dvo16a}, $\mathbf{J}_\mathrm{CME}$ can be
non-vanishing only if the mass of charged particles, forming the current,
is exactly equal to zero, i.e. the chiral symmetry is restored. For
the case of electrons the restoration of the chiral symmetry is unlikely
at reasonable densities which can be found in present universe~\cite{Rub86}.
The chiral symmetry can be unbroken in quark matter owing to the strong
interaction effects~\cite{BulCar16}. The magnetic fields generation
in quark matter, which can exist in some compact stars, was discussed
in Refs.~\cite{Dvo16b,Dvo16c}. Nevertheless this kind of situation
looks quite exotic.

Therefore the issue of the existence of an electric current $\mathbf{J}\sim\mathbf{B}$
for massive particles, which can lead to the magnetic field instability,
is quite important for the development of astrophysical magnetic fields models.
One of the example of such a current in electroweak matter was proposed
in Ref.~\cite{SemSok04}. However, the model developed in Ref.~\cite{SemSok04}
implies the inhomogeneity of background matter. This fact imposes
the restriction on the scale of the magnetic field generated.

In the present work, we discuss another scenario for the magnetic
field instability. It involves the consideration of the electroweak
interaction of massive fermions with background matter along with
nonzero anomalous magnetic moments of these fermions. Note that the
electroweak interaction implies the generic parity violation which
can provide the magnetic field instability.
%Recently, the interpretation
%of CME in terms of the effective magnetic moment was considered in
%Ref.~\cite{KhaSteYee17}.

This work is organized as follows. First, we discuss the Dirac equation
for a massive electron with a nonzero anomalous magnetic moment, electroweakly
interacting with background matter under the influence of an external
magnetic field. Using the previously obtained solution of this Dirac
equation, we calculate the electric current of these electrons along
the magnetic field direction. This current turns out to be nonzero.
Then we consider a particular situation of a strongly degenerate electron
gas, which can be found inside a neutron star (NS). Finally we apply
our results for the description of the amplification of the magnetic
field in NS and briefly discuss the implication of our findings to
explain the electromagnetic radiation of compact stars.

Let us consider a fermion (an electron) with the mass $m$ and the anomalous magnetic
moment $\mu$. This electron is taken to interact electroweakly with
nonmoving and unpolarized background matter, consisting of neutrons
and protons, under the influence of the external magnetic field along
the $z$-axis, $\mathbf{B}=B\mathbf{e}_{z}$. Accounting for the forward
scattering off background fermions in the Fermi approximation, the
Dirac equation for the electron has the form,
\begin{equation}\label{eq:Direq}
  \left[
    \gamma_{\mu}P^{\mu}-m-\mu B\Sigma_{3}-\gamma^{0}
    \left(
      V_{\mathrm{R}}
      P_{\mathrm{R}} +
      V_{\mathrm{L}}
      P_{\mathrm{L}}
    \right)
  \right]
  \psi = 0,
\end{equation}
where $P^{\mu}=\mathrm{i}\partial^{\mu}+eA^{\mu}$, $A^{\mu}=\left(0,0,Bx,0\right)$
is the vector potential, $e>0$ is the absolute value of the elementary
charge, $P_{\mathrm{R,L}} = (1\pm\gamma^{5})/2$ are the chiral projectors,
$\gamma^{\mu}= \left( \gamma^{0},\bm{\gamma} \right)$, $\gamma^{5}=\mathrm{i}\gamma^{0}\gamma^{1}\gamma^{2}\gamma^{3}$,
and $\Sigma_{3}=\gamma^{0}\gamma^{3}\gamma^{5}$ are the Dirac matrices. The effective potentials of the electroweak interaction $V_{\mathrm{R,L}}$
have the form~\cite{DvoSem15a},
\begin{equation}\label{eq:VRL}
  V_{\mathrm{R}}= -\frac{G_{\mathrm{F}}}{\sqrt{2}}
  \left[
    n_{n}-n_{p}(1-4\xi)
  \right]
  2\xi,
  \quad
%  \nonumber
%  \\
  V_{\mathrm{L}}= -\frac{G_{\mathrm{F}}}{\sqrt{2}}
  \left[
    n_{n}-n_{p}(1-4\xi)
  \right](2\xi-1),
\end{equation}
where $n_{n,p}$ are the number densities of neutrons and protons,
$G_{\mathrm{F}}=1.17\times10^{-5}\,\text{GeV}^{-2}$ is the Fermi
constant, and $\xi=\sin^{2}\theta_{\mathrm{W}}\approx0.23$ is the
Weinberg parameter.

The solution of Eq.~(\ref{eq:Direq}) has the form~\cite{BalStuTok12},
\begin{align}\label{eq:wavefun}
  \psi = & \exp
  \left(
    -\mathrm{i} Et +\mathrm{i} p_{y} y +\mathrm{i} p_{z} z
  \right)
  \times
  (C_{1} u_{\mathrm{n} -1}, \mathrm{i} C_{2} u_{\mathrm{n}},
  C_{3} u_{\mathrm{n} -1}, \mathrm{i} C_{4} u_{\mathrm{n}})^{\mathrm{T}},
  \quad
  \nonumber
  \\
  u_{\mathrm{n}} (\eta) = &
  \left(
    \frac{eB}{\pi}
  \right)^{1/4}
  \exp
  \left(
    -\frac{\eta^{2}}{2}
  \right)
  \frac{H_{\mathrm{n}} (\eta)}{\sqrt{2^{\mathrm{n}}\mathrm{n}!}},
  \quad
  \mathrm{n} = 0,1,\dotsc,
\end{align}
where $-\infty < p_{y, z} < +\infty$, $H_{\mathrm{n}}(\eta)$
are the Hermite polynomials, $\eta=\sqrt{eB}x + p_{y} / \sqrt{eB}$, and
$C_{i}$, with $i=1,\dots,4$, are the spin coefficients. For the definiteness,
we will use below the chiral representation for the Dirac matrices. It is
convenient to normalize the wave function $\psi$ as
\begin{equation}
  \int\mathrm{d}^{3}x
  \psi_{p_{y}p_{z}\mathrm{n}}^{\dagger}\psi_{p'_{y}p'_{z}\mathrm{n}'} =
  \delta\left(p_{y}-p'_{y}\right)\delta\left(p_{z}-p'_{z}\right)\delta_{\mathrm{nn}'},
\end{equation}
at any moment of time.

The energy levels $E$ for $\mathrm{n}>0$ have the form~\cite{BalStuTok12},
\begin{align}\label{eq:En}
  E = & \bar{V}+\mathcal{E},  
  \quad
  \mathcal{E}=\sqrt{p_{z}^{2}+m^{2}+2eB\mathrm{n} +
  \left(
    \mu B
  \right)^{2} + V_{5}^{2}+2sR^{2}},
  \notag
  \\
  R^{2} = & \sqrt{
  \left(
    p_{z} V_{5} - \mu B m
  \right)^{2} +
  2eB\mathrm{n}
  \left[
    V_{5}^{2} +
    \left(
      \mu B
    \right)^{2}
  \right]},
\end{align}
where $s=\pm1$ is the discrete spin quantum number, $\bar{V}=\left(V_{\mathrm{L}}+V_{\mathrm{R}}\right)/2$,
and $V_{5}=\left(V_{\mathrm{L}}-V_{\mathrm{R}}\right)/2$. At $\mathrm{n}=0$,
the energy spectrum reads
\begin{equation}\label{eq:E0}
  E = \bar{V} + \sqrt{
  \left(
    p_{z}+V_{5}
  \right)^{2} +
  \left(
    m-\mu B
  \right)^{2}}.
\end{equation}
It should be noted that, at lowest energy level, the electron spin
has only one direction since $C_{1}=C_{3}=0$. In Eqs.~(\ref{eq:En})
and~(\ref{eq:E0}), we present the solution only for particles (electrons) rather than for antiparticles (positrons).

Using the exact solution of the Dirac equation, we can calculate the
electric current of electrons in this matter. This current has the
form~\cite{Vil80},
\begin{equation}\label{eq:Jzgen}
  \mathbf{J} =
  - e
  \sum_{\mathrm{n}=0}^{\infty}
  \sum_{s}
  \int_{-\infty}^{+\infty}
  \mathrm{d}p_{y}\mathrm{d}p_{z}
  \bar{\psi}\bm{\gamma}\psi f(E-\chi),
\end{equation}
where $f(E)=\left[\exp(\beta E)+1\right]^{-1}$ is the Fermi-Dirac
distribution function, $\beta=1/T$ is the reciprocal temperature,
and $\chi$ is the chemical potential.

First, we notice that the transverse components of the electric current $J_{x,y}\sim\bar{\psi}\gamma^{1,2}\psi$ are vanishing 
because of the orthogonality of Hermite functions with different indexes. The contribution
of the lowest energy level with $\mathrm{n}=0$ to the electric current
along the magnetic field $J_{z}\sim\bar{\psi}\gamma^{3}\psi$ is also vanishing:
$J_{z}^{(\mathrm{n}=0)}=0$. This result is valid for arbitrary parameters
$m$, $\mu$, $V_{5}$, and $\chi$, and $B$.

The contributions of the higher energy levels with $\mathrm{n}>0$
to $J_{z}$ can be obtained using the expressions for the spin coefficients
$C_{i}$ also found in Ref.~\cite{BalStuTok12},
\begin{equation}\label{eq:Jzn>0}
  J_{z}^{(\mathrm{n}>0)} =
  -\frac{e^{2}B}{(2\pi)^{2}}
  \sum_{\mathrm{n}=1}^{\infty}
  \sum_{s=\pm1}
  \int_{-\infty}^{+\infty}
  \frac{\mathrm{d}p_{z}}{\mathcal{E}}
  \left[
    p_{z}
    \left(
      1+s\frac{V_{5}^{2}}{R^{2}}
    \right) -
    s\frac{\mu BmV_{5}}{R^{2}}
  \right]f(E-\chi).
\end{equation}
The first nonzero term in Eq.~(\ref{eq:Jzn>0}) is proportional to
$\mu B$ and $V_{5}$,
\begin{align}\label{eq:JznV5mu}
  J_{z}= & \mu mV_{5}B^{2}\frac{e^{2}}{\pi^{2}}
  \sum_{\mathrm{n}=1}^{\infty}
  \int_{0}^{+\infty}
  \frac{\mathrm{d}p}{\mathcal{E}_{\mathrm{eff}}^{2}}
  \left[
    \left(
      1-\frac{3p^{2}}{\mathcal{E}_{\mathrm{eff}}^{2}}
    \right)
    \left(
      f'-\frac{f}{\mathcal{E}_{\mathrm{eff}}}
    \right) +
    \frac{p^{2}}{\mathcal{E}_{\mathrm{eff}}}f''
  \right],
\end{align}
where $\mathcal{E}_{\mathrm{eff}}=\sqrt{p^{2}+m_{\mathrm{eff}}^{2}}$
and $m_{\mathrm{eff}}=\sqrt{m^{2}+2eB\mathrm{n}}$. The argument of the distribution
function in Eq.~(\ref{eq:JznV5mu}) is $\mathcal{E}_{\mathrm{eff}}+\bar{V}-\chi$.

Let us consider the case of a strongly degenerate electron gas. In
this situation, $f=\theta(\chi-\bar{V}-\mathcal{E}_{\mathrm{eff}})$,
where $\theta(z)$ is the Heaviside step function. We can also disregard
the positrons contribution to $J_{z}$. The direct calculation of
the current in Eq.~(\ref{eq:JznV5mu}) gives
\begin{equation}\label{eq:Jzdeggen}
  J_{z}= - 2\mu mV_{5}B^{2}
  \frac{e^{2}}{\pi^{2}
  \tilde{\chi}^{3}}
  \sum_{\mathrm{n}=1}^{\infty}
  \sqrt{\tilde{\chi}^{2}-m_{\mathrm{eff}}^{2}}\theta
  \left(
    \tilde{\chi}-m_{\mathrm{eff}}
  \right),
\end{equation}
where $\tilde{\chi}=\chi-\bar{V}$.

One can see that $J_{z}$ in Eq.~(\ref{eq:Jzdeggen})
is nonzero if $B<\tilde{B}$, where $\tilde{B}=\left(\tilde{\chi}^{2}-m^{2}\right)/2e$.
If the magnetic field is strong enough and is close to $\tilde{B}$,
then only the first energy level with $\mathrm{n}=1$ contributes
to $J_{z}$, giving one $J_{z} = - 8\mu mV_{5}B^{2} \alpha_{\mathrm{em}} \sqrt{\tilde{\chi}^{2}-m^{2}-2eB} / \pi\tilde{\chi}^{3} \to 0$,
where $\alpha_{\mathrm{em}}=e^{2}/4\pi \approx 7.3\times 10^{-3}$ is the fine structure constant. In the opposite situation,
when $B\ll\tilde{B}$, one gets that $J_{z} = - 8\alpha_{\mathrm{em}}\mu mV_{5}B \left( \tilde{\chi}^{2}-m^{2}-2eB \right)^{3/2} / 3\pi e \tilde{\chi}^{3} \approx -8\alpha_{\mathrm{em}}\mu mV_{5}B/3\pi e$,
i.e. the current is proportional to the magnetic field strength.

To study the evolution of the magnetic field in the presence of the current in Eq.~\eqref{eq:Jzdeggen} we return there to the vector notations,
\begin{equation}\label{eq:JPiB}
  \mathbf{J} = \Pi\mathbf{B},
  \quad
  \Pi=-8\mu mV_{5}B\frac{\alpha_{\mathrm{em}}}{\pi\tilde{\chi}^{3}}
  \sum_{\mathrm{n}=1}^{N}
  \sqrt{\tilde{\chi}^{2}-m_{\mathrm{eff}}^{2}},
\end{equation}
where $N$ is maximal integer, for which $\tilde{\chi}^{2}-m^{2}-2eBN\geq0$, and take into account the current in Eq.~\eqref{eq:JPiB} in the Maxwell equations along with the usual ohmic current $\mathbf{J}=\sigma_{\mathrm{cond}}\mathbf{E}$,
where $\sigma_{\mathrm{cond}}$ is the matter conductivity and $\mathbf{E}$
is the electric field.

Considering the magnetohydrodynamic approximation,
which reads $\sigma_{\mathrm{cond}}\gg\omega$, where $\omega$ is
the typical frequency of the electromagnetic fields variation, we
derive the modified Faraday equation for the magnetic field evolution,
\begin{equation}\label{eq:mFe}
  \frac{\partial\mathbf{B}}{\partial t} =
  \frac{1}{\sigma_{\mathrm{cond}}}\nabla \times
  \left(
    \Pi\mathbf{B}
  \right) +
  \frac{1}{\sigma_{\mathrm{cond}}}\nabla^{2}\mathbf{B},
\end{equation}
where we neglect the coordinate dependence of $\sigma_{\mathrm{cond}}$.

Let us consider the evolution of the magnetic field given by the Chern-Simons
wave, corresponding to the maximal negative magnetic 
helicity, $\mathbf{B}(z,t)=B(t)\left(\mathbf{e}_{x}\cos kz+\mathbf{e}_{y}\sin kz\right)$,
where $k=1/L$ is the wave number determining the length scale of
the magnetic field $L$ and $B(t)$ is the wave amplitude which can depend on time. In this situation we can
neglect the coordinate dependence of $\Pi$ in Eq.~(\ref{eq:mFe})
and the equation for $B$ takes the form,
\begin{equation}\label{eq:dotB}
  \dot{B}=-\frac{k}{\sigma_{\mathrm{cond}}}
  \left(
    k+\Pi
  \right)B.
\end{equation}
Since $\Pi$ in Eq.~(\ref{eq:JPiB}) is negative, the magnetic field
described by Eq.~(\ref{eq:dotB}) can be unstable since $\dot{B} > 0$.

We shall apply Eq.~(\ref{eq:JPiB}) to describe the magnetic field
amplification in a dense degenerate matter which can be found in NS.
In this situation, $n_{n}=1.8\times10^{38}\,\text{cm}^{-3}$ and $n_{p}\ll n_{n}$.
Using Eq.~(\ref{eq:VRL}), one
gets that $V_{5} = G_\mathrm{F} n_n /2\sqrt{2} = 6\,\text{eV}$. The number density of electrons can
reach several percent of the nucleon density in NS. We shall take
that $n_{e}=9\times10^{36}\,\text{cm}^{-3}$, which gives one $\chi = (3\pi^2 n_e)^{1/3} = 125\,\text{MeV}$~\cite{DvoSem15b}.
Thus electrons are ultrarelativistic and we can take that $\tilde{\chi}\approx\chi$.
We shall study the magnetic field evolution in NS in the time interval
$t_{0}<t<t_{\mathrm{max}}$, where $t_{0}\sim10^{2}\,\text{yr}$ and
$t_{\mathrm{max}}\sim10^{6}\,\text{yr}$. In this time interval, NS
cools down from $T_{0}\sim10^{8}\,\text{K}$ mainly by the neutrino
emission~\cite{YakPet04}. In this situation, the matter conductivity
in Eq.~(\ref{eq:dotB}) becomes time dependent $\sigma_{\mathrm{cond}}(t)=\sigma_{0}(t/t_{0})^{1/3}$~\cite{DvoSem15b},
where $\sigma_{0}=2.7\times10^{5}\,\text{GeV}$.

We shall discuss the amplification of the seed magnetic field $B_{0}=10^{12}\,\text{G}$, which is typical for a young pulsar.
In such strong magnetic fields, the anomalous magnetic moment of an
electron was found in Refs.~\cite{Ter69,Ter69a} to depend on the magnetic
field strength. We can approximate $\mu$ as
\begin{equation}\label{eq:mu}
  \mu = \frac{e}{2m}\frac{\alpha_{\mathrm{em}}}{2\pi}
  \left(
    1-\frac{B}{B_{c}}
  \right),
\end{equation}
where $B_{c}=m^{2}/e=4.4\times10^{13}\,\text{G}$. Note that Eq.~(\ref{eq:mu})
accounts for the change of the sign of $\mu$ at $B\approx B_{c}$
predicted in Ref.~\cite{Ter69}.

\begin{figure}
  \centering
  \subfigure[]
  {\label{1a}
  \includegraphics[scale=.12]{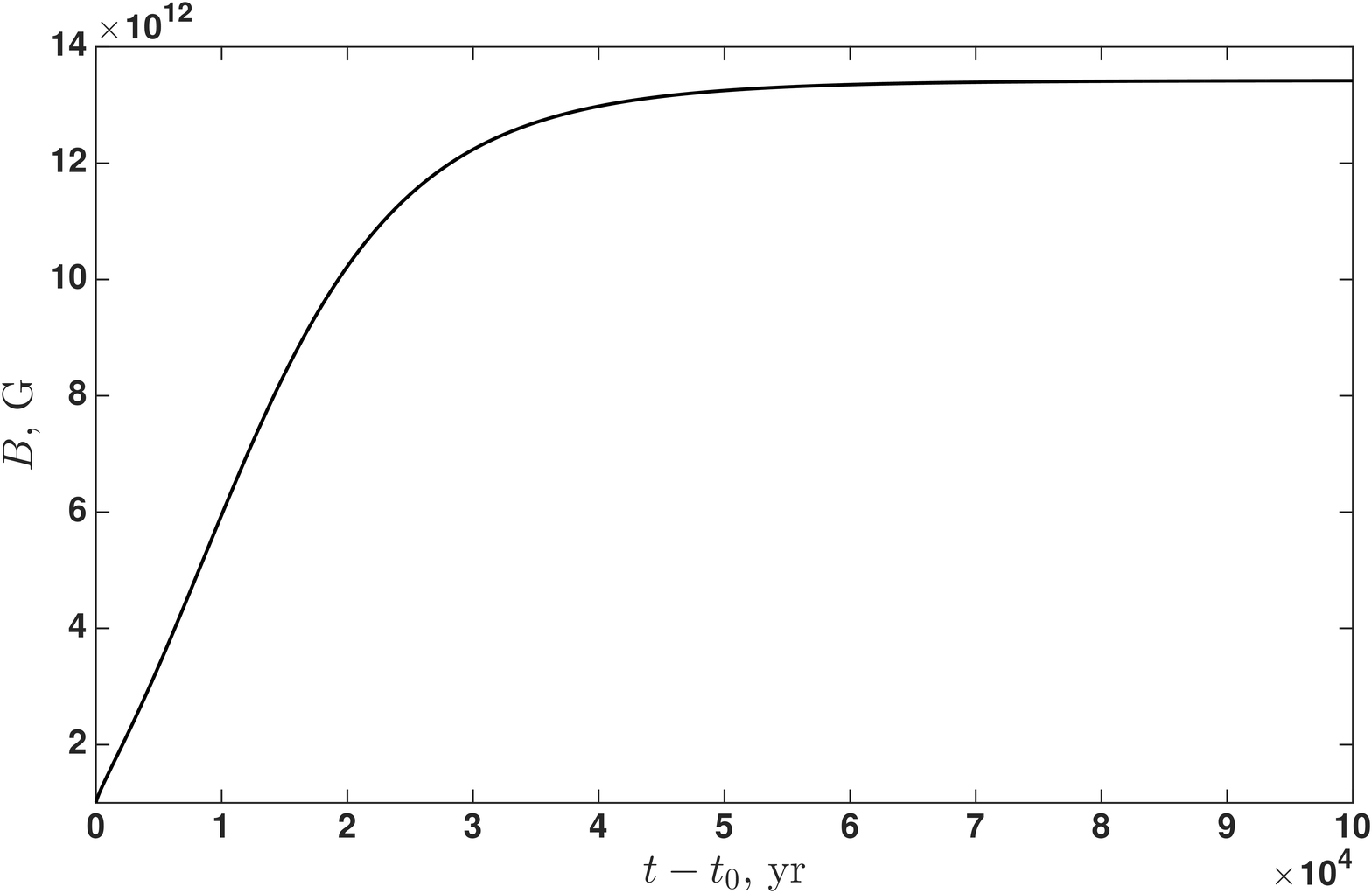}}
  \hskip-.7cm
  \subfigure[]
  {\label{1b}
  \includegraphics[scale=.12]{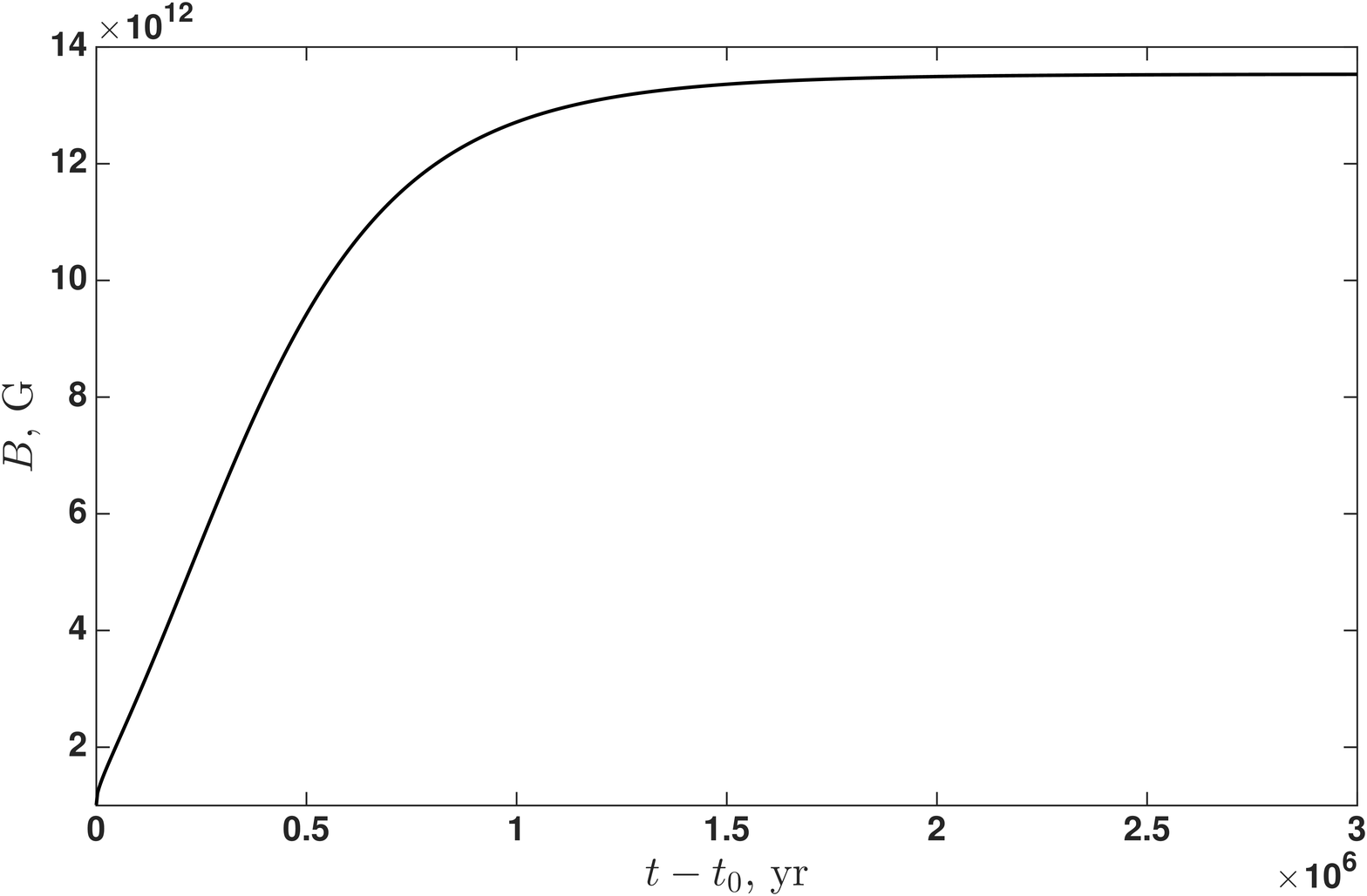}}
  \protect
  \caption{Magnetic field evolution obtained by the numerical solution of  
  Eq.~(\ref{eq:dotB}) for different length scales.
  (a) $L=10^{2}\,\text{cm}$, and (b) $L=10^{3}\,\text{cm}$.
  \label{fig:Bfield}}
\end{figure}

The evolution of the magnetic field for the chosen initial conditions
is shown in Fig.~\ref{fig:Bfield} for different length scales. One
can see that, if the magnetic
field is enhanced from $B_{0}=10^{12}\,\text{G}$, it reaches the saturated strength $B_{\mathrm{sat}} \approx 1.3\times10^{13}\,\text{G}$.
Thus, both quenching factors in Eqs.~(\ref{eq:JPiB}) and~(\ref{eq:mu})
are important. One can see in Fig.~\ref{fig:Bfield} that a larger
scale magnetic field grows slower. The further enhancement of the
magnetic field scale compared to $L=10^{3}\,\text{cm}$, shown in Fig.~\ref{1b}, is inexpedient since the growths time
would significantly exceed $10^{6}\,\text{yr}$. At such long evolution
times, NS cools down by the photon emission from the stellar surface
rather than by the neutrino emission~\cite{YakPet04}.

The energy source, powering the magnetic field growth shown in Fig.~\ref{fig:Bfield},
can be the kinetic energy of the stellar rotation. To describe the
energy transmission from the rotational motion of matter to the magnetic
field one should take into account the advection term $\nabla\left(\mathbf{v}\times\mathbf{B}\right)$
in the right hand side of Eq.~(\ref{eq:mFe}). Here $\mathbf{v}$
is the matter velocity.

Moreover one should assume the differential
rotation of NS~\cite{Sha00}.
%For this purpose we should take that
%NS is not in a superfluid state. This case is not excluded by the
%observational data~\cite{GneYakPot01}.
The NS spin-down because of the magnetic field enhancement can be estimated basing on the conservation of the total energy of a star: $I\Omega^{2}/2+B^{2}V/2 = \text{const}$, where $I$
is the moment of inertia of NS, $\Omega$ is the angular velocity,
and $V$ is the NS volume.

Taking the NS radius $R\sim10\,\text{km}$ and the initial rotation
period $P_{0}\sim10^{-3}\,\text{s}$, we get for  $B_{\mathrm{sat}}\approx1.3\times10^{13}\,\text{G}$,
shown in Fig.~\ref{fig:Bfield}, that the relative change of the period
is $(P-P_{0})/P_{0}\sim10^{-9}$. Hence only a small fraction of the initial rotational energy is transmitted to the energy of a growing magnetic field.

The obtained results can be used for the explanation of electromagnetic
flashes emitted by magnetars within the recently proposed model of a thermoplastic wave~\cite{BelLev14}, which can be excited by small-scale, with $L \sim (10^2 - 10^3)\,\text{cm}$, fluctuations of the magnetic field having the strength $B \gtrsim 10^{13}\,\text{G}$~\cite{Dvo16c}. The evolution of fields with such characteristics is shown in Fig.~\ref{fig:Bfield}.

In conclusion it is interesting to compare the appearance of the new current along the magnetic field in Eq.~\eqref{eq:JznV5mu} with CME~\cite{Vil80}, which is known to get the contribution only from massless electrons at the zero Landau level in an external magnetic field. Since left electrons move along the magnetic field and right particles move in the opposite direction, the current in Eq.~\eqref{eq:CME} is nonzero until there are different populations of the zero Landau level by left and right particles, i.e. $\mu_\mathrm{R}\neq\mu_\mathrm{L}$. Electrons at higher Landau levels can move arbitrarily with respect to the magnetic field. Therefore, CME is caused by an asymmetric motion of massless particles along the external magnetic field.

If we consider massive electrons with nonzero anomalous magnetic moments, electroweakly interacting with background matter, then, unlike CME, the motion of such particles at the lowest energy level with $\mathrm{n} = 0$ is symmetric with respect to  the magnetic field, i.e. $-\infty < p_z < +\infty$ for them (one can see it if we replace $p_z \to p_z - V_5$ in Eq.~\eqref{eq:E0}). On the contrary, higher energy levels with $\mathrm{n} > 0$ in Eq.~\eqref{eq:En} are not symmetric with respect to the transformation $p_z \to - p_z$. The reflectional symmetry cannot be restored by any replacement of $p_z$. Therefore electrons having $p_z > 0$ and $p_z < 0$ will have different energies and hence different velocities $v_z = p_z / \mathcal{E}$. Thus $J_z \sim \langle v_z \rangle \neq 0$, with only higher energy levels contributing to it. It is interesting to mention that the term in Eq.~\eqref{eq:En}, which violates the reflectional symmetry $p_z \to - p_z$, is proportional to $\mu B m V_5$. It is this factor which $J_z$ in Eq.~\eqref{eq:JznV5mu} is proportional to. Thus, one can see that the nonzero current $\mathbf{J} \sim \mathbf{B}$ results from the asymmetric motion of particles along $\mathbf{B}$. This asymmetry is caused by the simultaneous presence of three factors: nonzero $m$ and $\mu$, as well as the electroweak interaction with background matter $\sim V_5$.

It should be noted that, in addition to the electroweak interaction between electrons and background fermions, taken into account in Eqs.~\eqref{eq:Direq} and~\eqref{eq:VRL}, and leading to an electric current $\mathbf{J} \sim \mathbf{B}$ in Eq.~\eqref{eq:JPiB}, electrons also interact electromagnetically with background protons and neutrons. Considering, for definiteness, the electromagnetic interaction between electrons and a homogeneous gas of non-moving protons with a constant density, we find that the following additional term appears in the left hand side of Eq.~\eqref{eq:Direq}: $\sim [\dotsb + e^2 \gamma^0 f_0] \psi$, where $f_0\sim n_p / \omega_p^2$ is a quantity proportional to the zero component of the proton current and $\omega_p$ is the plasma frequency in the considered matter.

The strength of the electromagnetic interaction is much higher than that of the electroweak interaction, $e^2 f_0 \gg G_\mathrm{F} n_p$, since, in the degenerate matter, one has $\omega_p^2\sim \alpha_\mathrm{em} \chi^2$~\cite{BraSeg93} and $e^2 f_0\sim 10^{2}\,\text{GeV}^{-2} \times n_p$ for $ \chi\sim 10^2\,\text{MeV}$ (see above). Nevertheless, the additional contribution of the electromagnetic interaction in Eq.~\eqref{eq:Direq} can be eliminated by the gauge transformation $\psi\to \psi'= \exp (-ie^{2} f_{0} t) \psi$ in case of matter with constant density, $f_0\sim n_p = \text{const}$. The contribution of the electroweak interaction in Eq.~\eqref{eq:Direq}, $\gamma^0 (V_{\mathrm{R}} P_{\mathrm{R}} + V_{\mathrm{L}} P_{\mathrm{ L}}) \to \gamma^0 \gamma^5 V_5$ can not be eliminated by any gauge transformation because of the presence of the matrix $\gamma^5$ indicating the parity violation in electroweak interactions. Similarly, it can be shown that the electromagnetic interaction between electrons and neutrons, due to the presence of the magnetic form factor of neutrons~\cite{PerPunVan07}, does not give rise to a current $\mathbf{J} \sim \mathbf{B}$ in the case of a homogeneous, unpolarized neutron matter with a constant density.

The necessity of the presence of the contribution of a parity violating interaction in the generation of the current $\mathbf{J} = \Pi \mathbf{B}$ in Eq.~\eqref{eq:JPiB} in the system of massive fermions follows also from the fact that the parameter $\Pi$ should be a pseudoscalar. Electromagnetic interaction is known to be parity conserving. That is why it does not contribute to $\Pi$ in the system in question.

\section*{Acknowledgements}

I am thankful to the Tomsk State University Competitiveness Improvement
Program and RFBR (research project No.~15-02-00293) for a partial support.

\end{document}